\documentclass[11pt,a4paper]{article}
\usepackage{jheppub} % for details on the use of the package, please
\def\eq#1\en{\begin{equation}#1\end{equation}}  
\def\eqa#1\ena{\begin{align}#1\end{align}}
\def\eqg#1\eng{\begin{gather}#1\end{gather}}
\newcommand{\lb}[1]{\label{e:#1}}
\newcommand{\rlb}[1]{\eqref{e:#1}} 
\newcommand{\nl}{\notag\\}

\newcommand{\sumtwo}[2]%
{\mathop{\sum_{#1}}_{#2}}
\newcommand{\prodtwo}[2]%
{\mathop{\prod_{#1}}_{#2}}
\newcommand{\mintwo}[2]%
{\mathop{\min_{#1}}_{#2}}
\newcommand{\ket}[1]{|#1\rangle}
\newcommand{\bra}[1]{\langle#1|}
\newcommand{\braket}[1]{\langle #1\rangle}
\newcommand{\calF}{\mathcal{F}}
\newcommand{\tF}{\tilde{\mathcal{F}}}
\newcommand{\calH}{\mathcal{H}}
\newcommand{\calS}{\mathcal{S}}
\newcommand{\calL}{\mathcal{L}}
\newcommand{\bb}{\bar{\ell}}
\newcommand{\bell}{\bar{\ell}}
\newcommand{\bV}{\bar{V}}
\newcommand{\dV}{\partial V}
\newcommand{\dbV}{\partial\bV}
\newcommand{\bsf}{\boldsymbol{f}}

\newcommand{\bk}{\boldsymbol{k}}
\newcommand{\bp}{\boldsymbol{p}}
\newcommand{\bm}{\boldsymbol{m}}

\newcommand{\oN}{\{0,1,\ldots,N-1\}}
\newcommand{\eL}{{\ell\in\calL}}
\newcommand{\hrho}{\hat{\rho}}
\newcommand{\set}[2]{\bigl\{\,#1\,\bigl|\,\text{#2}\,\bigr\}}

\newcommand{\nn}{\nonumber}

\newcommand{\pslash}{p\kern-1ex /}
\newcommand{\qslash}{q\kern-1ex /}
\newcommand{\lslash}{l\kern-1ex /}
\newcommand{\sslash}{s\kern-1ex /}
\newcommand{\kaslash}{k_a\kern-2ex /}
\newcommand{\kbslash}{k_b\kern-2ex /}
\newcommand{\Dslash}{{\cal D}\kern-1.5ex /}

\newcommand{\tr}{{\rm tr}}

\newcommand{\beqa}{\begin{eqnarray}}
\newcommand{\eeqa}{\end{eqnarray}}

\def\Tr{\text{Tr}}
\def\tr{\text{tr}}

\newcommand{\be}{\begin{equation}}
\newcommand{\ba}{\begin{eqnarray}}
\newcommand{\ea}{\end{eqnarray}}
\newcommand{\ee}{\end{equation}}

\title{On the definition of entanglement entropy in lattice gauge theories}

\author[a]{Sinya Aoki,}
\author[a]{Takumi Iritani,}
\author[a]{Masahiro Nozaki,}
\author[a]{Tokiro Numasawa,}
\author[a]{Noburo Shiba,}
\author[b]{Hal Tasaki}

\affiliation[a]{Yukawa Institute for Theoretical Physics, Kyoto University, \\
Kitashirakawa Oiwakechou, Sakyo-ku, Kyoto 606-8502, Japan}
\affiliation[b]{Department of Physics, Gakushuin University, \\
Mejiro 1-5-1, Toshima, Tokyo 171-8588, Japan}

\emailAdd{saoki@yukawa.kyoto-u.ac.jp}
\emailAdd{iritani@yukawa.kyoto-u.ac.jp}
\emailAdd{mnozaki@yukawa.kyoto-u.ac.jp}
\emailAdd{numasawa@yukawa.kyoto-u.ac.jp}
\emailAdd{shibn@yukawa.kyoto-u.ac.jp}
\emailAdd{hal.tasaki@gakushuin.ac.jp}

\abstract{
We focus on the issue of proper definition of entanglement entropy in lattice gauge theories, and examine a naive definition where gauge invariant states are viewed as elements of an extended Hilbert space which contains gauge non-invariant states as well.  Working in the extended Hilbert space, we can define entanglement entropy associated with an arbitrary subset of links, not only for abelian but also for non-abelian theories.  We then derive the associated replica formula.  We also discuss the issue of gauge invariance of the entanglement entropy.  
In the $Z_N$ gauge theories in arbitrary space dimensions, we show that all the standard properties 
of the entanglement entropy, e.g. the strong subadditivity, 
hold in our definition. 
We study the entanglement entropy for special states, including  
the topological states for the $Z_N$ gauge theories in arbitrary dimensions.
We discuss  relations of our definition to other proposals.

}


\subheader{\hfill YITP-2015-8}

\begin{document} 

\maketitle

\section{Introduction}
\label{intro}
Entanglement entropy plays important roles in various fields of quantum physics including string theory\cite{t4, t1, sw1, sw2, r, t2, t3, t5}, condensed matter physics\cite{wen, pre, mul, hal, wen2, gon2}, and the physics of the black hole \cite{sus, kab1, kab2, sh1, sh2}. 
It is believed that entanglement entropy characterizes various aspects of quantum states in a simple and unified manner.

In the context of lattice gauge theories, entanglement entropy is expected to be a useful tool for studying confinement / deconfinement transitions (or crossover) \cite{NT, dcp, lv}.  It has been pointed out, however, that there is a subtle problem in the definition of entanglement entropy in gauge theories\cite{CHR2013, Radicevic2014, Donnelly2012, Don}.
When we calculate the entanglement entropy of a region $V$,  
we first express the Hilbert space of the total system 
as a tensor product of the Hilbert spaces of $V$ 
and that of $\bV$,  the complement of $V$.
Thus we trace out the degrees of freedom of  $\bar{V}$ 
and obtain the reduced density matrix of $V$. 
For gauge theories, however, the physical gauge invariant Hilbert space 
can not be factorized into a tensor product of the gauge invariant subspaces 
of $V$ and that of $\bar{V}$ 
due to the local gauge invariance at the boundary $\dV$ between $V$ and $\bV$.
This reflects the fact that the fundamental physical degrees of freedom contain 
Wilson loops, which are nonlocal operators.  
Due to the absence of the factorization into a tensor product, 
it is not straightforward to define the reduced density matrix 
of some region and to calculate the entanglement entropy. 
We need to specify the prescription to obtain the reduced density matrix 
of the region. 

In this paper, we propose a definition of the entanglement entropy 
in lattice gauge theories. 
We extend the gauge invariant Hilbert space to a larger Hilbert space in order to admit 
 the factorization into a tensor product of 
the gauge invariant subspaces 
of the region $V$ and the region $\bar{V}$ in this larger Hilbert space. 
The natural candidate of this larger Hilbert space is 
the whole (gauge non-invariant) Hilbert space of the link variables. 
We then obtain the reduced density matrix of the region $V$ 
by tracing out the link variables of the region $\bar{V}$. 
We define the entanglement entropy as the von Neumann entropy 
of the above reduced density matrix. 
We can define the entanglement entropy for an arbitrary subset of links. 
This definition is applicable not only for abelian theories but also for non-abelian ones. 
We then derive the replica formula to calculate the entanglement entropy in our definition. 

In the $Z_N$ gauge theories in arbitrary space dimensions, 
we express the whole Hilbert space by useful basis states, which are eigenstates of 
the gauge transformations \cite{Radicevic2014}. 
We argue that all the standard properties of entanglement entropy, e.g. the strong subadditivity, 
hold in our definition. 
We study the one for some special states. 
In particular, we calculate rigorously the one for the topological states in arbitrary space dimensions. 
We discuss  relations of our definition to other proposals. 
We also demonstrate that the entanglement entropy depends on the choice of the gauge fixing for some simple cases.
This indicates that 
one should not fix the gauge, at least on the boundary points between two regions, to calculate the entanglement entropy in gauge theories.

The present paper is organized as follows. 
In section 2, we give precise definitions of the geometry of our lattice 
and define the entanglement entropy. 
We discuss the gauge invariance of the reduced density matrix.
We also derive the replica formula here. 
In section 3, we consider the $Z_N$ gauge theories. 
We express the whole Hilbert space by eigenstates of 
the gauge transformations, and
derive an explicit expression of the entanglement entropy.  
We then argue that all the standard properties of entanglement entropy, e.g. the strong subadditivity, 
hold in our definition. 
We study the one for some special states. 
In particular, we calculate  the one for the topological states in arbitrary space dimensions. 
We discuss  relations of our definition to other proposals. 
In section 4, we summarize our investigations.
Some properties  of the $Z_N$ gauge theories used in the main text are given in appendix \ref{app:dim_f},
while gauge invariant states in non-abelian gauge theories are briefly discussed in \ref{app:general}.

\section{Naive definition of entanglement entropy in lattice gauge theories}
\label{def}
\subsection{Definition and some properties}
\paragraph*{Geometry}
We can treat quite general geometries and boundary conditions.

Our lattice is $(\calS,\calL)$, where $\calS$ denotes the set of sites $x,y,\ldots\in\calS$, and $\calL\subset\calS\times\calS$ the set of links.
We understand that $\ell=(x,y)$ and $\bell=(y,x)$ are different ways of expressing a single link in $\calL$.  This in particular means that $\ell\in V$ implies $\bb\in V$ for any subset $V\subset\calL$.
Here we do not assume a particular structure of our lattice such as regularity, so that a random lattice could be treated. Note that this setup can treat both periodic  and free boundary conditions for the whole lattice.

We define the boundary of a subset $V\subset\calL$ as 
\eq
\dV:=\set{x\in\calS}
{$(x,y)\in V$ and $(x,z)\in \bV$ for some $y,z\in\calS$},
\en
which is the set of sites in both $V$ and its complement $\bV=\calL\backslash V$.
Note also that $\dV=\partial\bV$.

\paragraph*{Naive definition of entanglement entropy}
We consider the global density matrix $\rho$  for gauge theories, whose elements are denoted  by
\beqa
\langle U \vert \rho \vert U^\prime \rangle &\equiv &  \rho(U;U^\prime) =\rho(U_V, U_{\bar V}; U_V^\prime, U_{\bar V}^\prime )
\eeqa
where $U$ represents a gauge configuration (a set of all link variables), $U =\{ U_\ell \ \vert\ \ell \in \calL \}$, while 
$U_V, U_{\bar V}$ are gauge configurations on $V$ and $\bar V$, 
$U_V =\{ U_\ell \ \vert\ \ell \in V \}$ and  $U_V =\{ U_\ell\ \vert\ \ell \in \bV \}$, respectively.

We propose to  define  a reduced density matrix  as
\beqa
\langle U_V \vert \rho_V \vert U_V^\prime \rangle \equiv \rho_V(U_V;U^\prime_V) &=& \int {\cal D}U_{\bar V} \, \rho(U_V, U_{\bar V}; U_V^\prime, U_{\bar V}),
\label{eq:rho_V}
\eeqa 
where ${\cal D}U_{\bV}$ denotes  a product of the group invariant integrals or sums. For the compact group,
we have ${\cal D} U_{\bV} = \prod_{\ell \in \bV} d U_\ell$, where $ d U_\ell$ is the Haar measure for the link variable $U_\ell$. 

The above definition of the reduced density matrix is a simple generalization of the reduced density matrix in spin systems, where the whole Hilbert space is a direct product of those of region $V$ and region $\bV$, ${\cal H} = {\cal H}_V \otimes  {\cal H}_{\bV}$.
In the case of gauge theories, on the other hand, due to the local gauge invariance, the gauge invariant full Hilbert space can not be factorized into a product of gauge invariant subspaces,  ${\cal H}_{\calL}^{\rm inv} \not= {\cal H}_V^{\rm inv} \otimes  {\cal H}_{\bV}^{\rm inv}$. Therefore the above reduced density matrix $\rho_V$ can not be obtained from a single  partial trace of $\rho$ over the gauge invariant subspace ${\cal H}_{\bV}^{\rm inv}$.
Without gauge invariance, however, the whole Hilbert space can be factorized as ${\cal H}_{\calL} = {\cal H}_V \otimes  {\cal H}_{\bV}$, so that
our definition of $\rho_V$ above can be understood as the partial trace of $\rho$  over the gauge non-invariant subspace ${\cal H}_{\bV}$ on $U_{\bV}$.
In the next section, we explicitly construct the reduced density matrix for the $Z_N$ gauge theories in an arbitrary dimensions, and explicitly construct an extension of $\rho_V$  to ${\cal H}_{\calL} = {\cal H}_V \otimes  {\cal H}_{\bV}$.

From the reduced density matrix, the entanglement entropy  can thus be defined as
\beqa
S(V) &=& - \tr \, [ \rho_V \log \rho_V] ,
\label{eq:EE}
\eeqa
where the trace is taken over ${\cal H}_V$.
The definitions (\ref{eq:rho_V}) and  (\ref{eq:EE}) are so simple that they can be used 
not only for discrete abelian theories but also for continuous 
non-abelian gauge theories without practical difficulties.

In the next section, we will see that this trace is reduced to a sum of traces in the gauge invariant subspace ${\cal H}_V^{\rm inv} \subset {\cal H}_V$ and 
discuss that
a basic properties such as the symmetric property and the strong subadditivity are satisfied for the $Z_N$ gauge theories.

\paragraph*{Gauge invariance}
In gauge theories, the global density matrix is gauge invariant as
\beqa
\rho(U^g;{U^\prime}^h) = \rho(U;U^\prime),
\eeqa
where the gauge transformation of the link variable $U_{\ell}$ is given by $U_{\ell}^g = g_x U_{\ell}g_{y}^\dagger$ with $\ell=(x,y)$.

On the other hand, the reduced density matrix $\rho_V$ in ~(\ref{eq:rho_V}) does not have such a gauge invariance. 
Indeed,
\beqa
\rho_V(U_V^{g_V};{U^\prime_V}^{h_V}) &=&  \int {\cal D}U_{\bar V} \, \rho(U_V^{g_V}, U_{\bar V}; {U_V^\prime}^{h_V}, U_{\bar V}) = \int {\cal D}U_{\bar V}^{g_{\bV}} \, \rho(U_V^{g_V}, U_{\bar V}^{g_{\bV}}; {U_V^\prime}^{h_V}, U_{\bar V}^{g_{\bV}}) \nonumber \\
&=& \int {\cal D}U_{\bar V} \, \rho(U_V, U_{\bar V}; (U_V^\prime)^{h_Vg_V^{-1}}, U_{\bar V})
 = \rho_V(U_V;(U^\prime_V)^{h_V g_V^{-1}}) ,
\eeqa 
where  the invariance of the measure $ {\cal D}U_{\bar V}^{g_{\bV}} = {\cal D}U_{\bar V} $ and
the gauge invariance of the full density matrix $\rho$ such as
\beqa
\rho(U_V^{g_V}, U_{\bV}^{g_{\bV}}; {U_V^\prime}^{g_V}, U_{\bV}^{g_{\bV}}) &=& \rho(U_V, U_{\bV}; U_V^\prime, U_{\bV})  
\eeqa
are used.
Therefore $\rho_V$ is invariant under diagonal gauge transformations ($g_V=h_V$) only.

This suggests that the reduced density matrix and thus its entanglement entropy may depend on the choice of the gauge if the gauge fixing is employed in the calculation.
Indeed, we will show in the next section that values of the entanglement entropy  are different for different  gauge fixing conditions in some simple cases for the $Z_N$ gauge theories. 
Because of this problem, it is important and sensible to calculate the entanglement entropy in the gauge invariant way without gauge fixing.

\subsection{Replica formula}
We briefly consider the replica formula for the entanglement entropy of lattice gauge theories based on our definition.

\paragraph*{Transfer matrix and path integral} 
In lattice gauge theories, the evolution  in a discrete time is given by the transfer matrix $\hat T$ (for example, see Refs.~\cite{Creutz1977, Luescher1977}), which is given by
\be
T(U, U^\prime):= \braket{U\vert \hat T \vert U^\prime} = \exp[ \frac{1}{2}S_{d} (U) ]\exp[ S_0(U,U^\prime) ] \exp[\frac{1}{2}S_{d}(U^\prime)]
\ee 
where
\ba
S_{d}(U)&=&\frac{1}{2g^2} \sum_{x\in \calS} \sumtwo{\mu=1,\nu=1}{\mu \not= \nu}^d \Tr\, \left[ U_{x,\mu} U_{x+\hat\mu,\nu} U_{x+\hat\nu,\mu}^\dagger U_{x,\nu}^\dagger\right] 
:= \frac{1}{2g^2} \sum_{x\in \calS} \sumtwo{\mu=1,\nu=1}{\mu \not= \nu}^d P_{\mu\nu}(x),
\\
S_0(U,U^\prime) &=& \frac{1}{2g^2} \sum_{x\in \calS}\sum_{\mu=1}^d \Tr \left[ U_{x,\mu} (U^\prime_{x,\mu})^\dagger + U_{x,\mu}^\prime (U_{x,\mu})^\dagger \right]
\ea
for the plaquette action on a $d$-dimensional hyper-cubic lattice with the coupling constant $g$.
We here define $P_{\mu\nu} (x)$ as a trace of the plaqiutte on $\mu\nu$ plane at $x$, and
 use the notation that $U_{x,\mu}:= U_{\ell}$ with $\ell=(x,x+\hat\mu)$ and $\hat\mu$ is an unit vector in the $\mu$ direction.  

The wave function for the vacuum state is obtained as
\ba
\lim_{N_T\rightarrow\infty}\bra{U}\hat P  (\hat T)^{N_T} \ket{\Psi} &=&\braket{U\vert 0} \braket{0\vert \Psi}
\ea
for an arbitrary gauge invariant state $\ket{\Psi}$ which satisfies $ \braket{0 \vert\Psi}\not=0$, where
$\hat P$ is a projection to the physical (gauge invariant) Hilbert space as
\be
\hat P :=  \prod_{x\in\calS} \int dg_x\, \hat E_x(g_x) .
\ee
Here $\hat E_x(g_x)$ generates the gauge transformation at $x$ by $g_x$.  Note that  $(\hat P)^2 =\hat P$ and $[\hat T, \hat P]=0$. While we explicitly write $\hat P$ in the above expression since $\bra{U}$ is not gauge invariant,  the formula without $\hat P$ is equaly correct since $\hat P\ket{\Psi} =\ket{\Psi}$.

Thus we can write
\ba
\bra{U} \hat P (\hat T)^{N_T} \ket{U^\prime} &=& \exp\left[\frac{1}{2} S_d( U )\right] 
\int_{U[0]=U^\prime}^{U[N_T]=U} {\cal D}U[t] 
\exp\left[ S_G(U[t]) \right] \exp\left[\frac{1}{2} S_d( U^\prime)\right]  
\lb{eq:path-integral}
\ea
where
\ba
 {\cal D}U[t] &:=&\prod_{t=1}^{N_T-1}\prod_{\ell\in\calL} dU_{\ell}[t] \prod_{t=0}^{N_T}\prod_{x\in\calS} dg_x[t] \\
  S_G(U[t]) &:=& \sum_{t=1}^{N_T-1} S_d( U[t] ) + \sum_{t=1}^{N_T} S_0(U^g[t-1], U^g[t])
  \lb{eq:action_G}
\ea
where $U_{x,\mu}^g[t] = g_x[t] U_{x,\mu}[t] g_{x+\hat\mu}^\dagger[t]$.   Defining a new gauge field as
$U_{z,0} := g_x^\dagger[t] g_x[t+1]$ where $z=(x,t)$ is a $d+1$ dimensional lattice point, and
introducing a new notation for gauge fields $U_{z,\mu}$ with $\mu=0,1,\cdots d$, we have
\ba
\bra{U_d} \hat P(\hat T)^{N_T} \ket{U_d^\prime} &=&  \exp\left[\frac{1}{2} S_d( U_d )\right] 
\int_{U_{z_0}=U_d^\prime}^{U_{z_T}=U_d} {\cal D}U \int {\cal D} g_{z_0} 
\exp\left[ S_{\rm plaq}(U_d) \right]
 \exp\left[\frac{1}{2} S_d( U^\prime)\right] \notag \\
\ea
where
\ba
S_{\rm plaq}(U) &=& \frac{1}{2g^2}\sum_{x\in\calS}\Bigl[\sum_{t=1}^{N_T-1}\sumtwo{\mu=0,\nu=0}{\mu\not=\nu}^d
P_{\mu\nu} (z) + \sum_{k=1}^d  \left\{P_{0k}(z_0)+P_{k0}(z_0)\right\}\Bigr], 
\ea
with $z_0=(x,0)$ and $z_T=(x,N_T)$. Here $U_{z_0}=U_d$ means $U_{z_0,k}=(U_d)_{x,k}$ for $k=1,2,\cdots, d$.
Note that since $S_{\rm plaq}$ and $S_d$ do not depend  on $g_{z_0}$, the gauge transformation left after the change of variables, we have
$\int {\cal D} g_{z_0} =1$ in the above expression.

\paragraph*{Path integral expression}
The (unnormalized) density matrix for the vacuum state, can be obtain as
\ba
\braket{U_d\vert \hat\rho \vert U_d^\prime}&=& 
e^{\frac{1}{2} [S_d( U_d )+S_d(U_d^\prime)]} 
\lim_{N_T\rightarrow\infty} \int{\cal D} U \delta(U_{z_0^-}-U_d)\delta(U_{z_0^+}-U_d^\prime) 
e^{S_{\rm plaq} (U)}\notag \\
&\times& 
e^{\frac{1}{2} [S_d( U_{z_T^+} )+S_d(U_{z_T^-})]} 
\ea 
where $x\in\calS$, $ -N_T \le t \le N_T$, $z_0^\pm =(x,0^\pm)$, and $z_T^\pm =(x,\pm N_T)$.

In practice, one often employs the periodic boundary condition at $\pm N_T$ in the Euclidian time, which correspond to the thermal density matrix  at temperature $T = 1/(2N_T a)$,  where $a$ is the lattice spacing. In this case, after interchanging $t=0$ and $t=\pm N_T$, we have
\ba
\braket{U_d\vert \hat\rho^T \vert U_d^\prime}&=& 
e^{\frac{1}{2} [S_d( U_d )+S_d(U_d^\prime)]} 
 \int_{U_{z_T^-}=U_d^\prime}^{U_{z_T^+}=U_d} {\cal D} U  e^{S_{\rm plaq}^T (U)}
\ea
where
\be
S^T_{\rm plaq}(U) = \frac{1}{2g^2}\sum_{x\in\calS}\Bigl[\sum_{t=-N_T+1}^{N_T-1}\sumtwo{\mu=0,\nu=0}{\mu\not=\nu}^d P_{\mu\nu}(z) + \sum_{k=1}^d\left\{ P_{0k}(z_T^-)+P_{k0}(z_T^-)\right\}\Bigr].
\ee
The density matrix for the vacuum state is reproduced from $\hat \rho^T$ by the $T\rightarrow 0$ limit.

\paragraph*{Reduced density matrix for replica formula}
We now consider two regions $V$ and $\bV =\calL \backslash V$, and denote $U=(U_V, U_{\bV})$ and
$U_d=(U_d{}_V, U_d{}_{\bV})$.
Then the reduced density matrix $\hat \rho_V^T$ can be written as
\ba
\rho_V^T(U_d{}_V; U_d^\prime{}_V):=\braket{U_d{}_V\vert \hat P\hat\rho_V^T \vert U_d^\prime{}_V}&=&  \int_{U_{z_T^-}{}_V=U_d^\prime{}_V}^{U_{z_T^+}{}_V=U_d{}_V} {\cal D} U  e^{\frac{1}{2} [S_d( U_d )+S_d(U_d^\prime)]}  e^{S_{\rm plaq}^T (U)} .
\lb{eq:path-rho}
\ea

The replica formula for the entanglement entropy in now given as
\ba
S(V) &=& \lim_{n\rightarrow 1} \frac{1}{1-n}\log\left( \frac{Z_{n}}{ Z_1^n }\right), \qquad
Z_n :=\Tr (\hat \rho_V^T)^n
\ea
where $Z_n$ can be expressed in the path-integral as
\ba
Z_n  &=&\left( \prod_{i=1}^n \int {\cal D}U_i\right)  \rho_V^T (U_1; U_2)  \rho_V^T (U_2; U_3)\cdots  \rho_V^T (U_n; U_1)  
\lb{eq:replica}
\ea
and $\rho_V^T (U_i,U_j)$ is given in \rlb{eq:path-rho}.
Note that \rlb{eq:replica} is invariant under the local gauge transformation $g$ in $d+1$ dimensions with the period $2N_T$ ( not $ 2n N_T$) at the boundary,
which satisfies 
\ba
 \qquad  g_{(x, 2k N_T)} = g_{(x, 0)} \quad \rm{ at }\ x\in \dV, \qquad k= 1,2,\cdots, n .
\ea

\section{$Z_N$ gauge theories in an arbitrary dimension}
\label{Z_N}
We consider the $Z_N$ gauge theories in this section.

\subsection{Some properties of divergence-free flux-configurations}
\paragraph*{Flux-configuration}
For each link $\ell\in\calL$, we associate a flux $k_\ell\in\oN$.
We assume the consistency $k_\ell=-k_{\bell}$.
Here and throughout the present paper, equalities for the flux are with respect to mod $N$.
We denote by $\bk=(k_\ell)_{\eL}$ a configuration of flux over the whole lattice which 
satisfies
$
d_x^{\calL}(\bk) = 0
$
at ${}^\forall x \in \calS$, where 
\eq
d_x^\calL (\bk):=\sumtwo{y\in\calS}{{\rm s.t.}\,(x,y)\in \calL}k_{(x,y)}.
\lb{div}
\en
is  the divergence of $\bk$ at $x$ associated with region $\calL$.  
We denote the set of all divergent-free $\bk$'s by $\tF$. 

Take an arbitrary subset $V\subset\calL$.
For any $\bk\in\tF$, let $R_V(\bk)$ be the  configuration obtained by omitting all the flux outside $V$.
We then denote by $\tF_V$ the set of $\bk'$ which is written as $\bk'=R_V(\bk)$ for some (not necessarily unique) $\bk\in\tF$.

\paragraph*{Incoming flux and decomposition of $\tF$}
Fix  an arbitrary subset  $V\subset\calL$.
For any $\bk\in\tF$, we define
\eq
\bsf_V(\bk):=\bigl(d_x^V(\bk)\bigr)_{x\in\dV},
\en
where $d_x^V$ is the divergence associated with the region $V$, obtained  by replacing $\calL \to V$ in \rlb{div}. 
Note that $\bsf_V(\bk)$ is the list of incoming flux at each site on the boundary $\dV$.
Recalling that $\dV=\partial\bV$ for $\bV=\calL\backslash V$, we have
\eq
\bsf_V(\bk)=-\bsf_{\bV}(\bk),
\en
which represents the conservation of flux at the boundary sites.

For a give subset  $V\subset\calL$, we say that $\bsf\in\oN^{\dV}$ is admissible if there exists at least one $\bk\in\tF$ such that $\bsf_V(\bk)=\bsf$.
Then we have a natural decomposition
\eq
\tF=\bigcup_{\bsf}\tF^{(\bsf)},
\lb{Fdec}
\en
where the union is over all admissible $\bsf$, and  
\eq
\tF^{(\bsf)}:=\set{\bk\in\tF}{$\bsf_V(\bk)=\bsf$}.
\en

It is remarkable that all $\tF^{(\bsf)}$ with admissible $\bsf$ are completely isomorphic to each other.
To see this, take arbitrary $\bsf_1$ and $\bsf_2$ which are admissible.
Choose and fix $\bk_1,\bk_2\in\tF$ such that $\bsf_V(\bk_i)=\bsf_i$ for $i=1,2$.
Then we define a map $\varphi_{1,2}:\tF^{(\bsf_1)}\to\tF^{(\bsf_2)}$ by $\varphi_{1,2}(\bk):=\bk-\bk_1+\bk_2$ and its inverse map $\varphi_{2,1}:\tF^{(\bsf_2)}\to\tF^{(\bsf_1)}$ by $\varphi_{2,1}(\bk):=\bk-\bk_2+\bk_1$. Since $\varphi_{1,2}(\bk_a) \not=  \varphi_{1,2}(\bk_b)$ for $\bk_a \not=\bk_b$, $\bk_{a,b}\in \tF^{(\bsf_1)}$ and a similar property for $\varphi_{2,1}$, these maps establish a one-to-one correspondence between the elements of $\tF^{(\bsf_1)}$ and $\tF^{(\bsf_2)}$: $\tF^{(\bsf_1)}$ and $\tF^{(\bsf_2)}$ are  isomorphic to each other.

Finally let us evaluate the number of all the admissible $\bsf$'s.
Decompose $V$ and $\bar V$  into connected components as $V=V_1\cup\cdots\cup V_n$ and
$\bV=\bar V_1\cup\cdots\cup \bar V_m$. (For example, see Fig.~\ref{f:connection} in appendix~\ref{app:dim_f}.) 
Correspondingly, the boundary $\dV$ is decomposed as $\dV=\dV_1\cup\cdots\cup\dV_n =\dbV= \dbV_1\cup\cdots\cup\dbV_m$.
Then the divergence-free condition for $\bk$ implies that an admissible incoming flux $\bsf=(f_x)_{x\in\dV}$ satisfies 
\beqa
\sum_{x\in\dV_i}f_x=0,  && i =1,\ldots,  n, \\
\sum_{x\in\dbV_j}f_x=0, && j=1,\ldots, m ,
\lb{ccond}
\eeqa
with an additional condition that 
\beqa
\sum_{i=1}^n \sum_{x\in\dV_i}f_x = \sum_{j=1}^m \sum_{x\in\dbV_j}f_x
\eeqa
for an arbitrary $\bsf$ even without satisfying the divergent-free condition.
Thus the total number of the admissible $\bsf$'s is readily found to be
$N^{|\dV|-(n+m-1)}$, where $|\dV|$ denotes the number of sites in $\dV$.
See appendix~\ref{app:dim_f} for a more rigorous discussion.

\paragraph*{Decomposition of $\bk$}
Let $V\subset\calL$ be a subset, and $\bsf$ be an admissible incoming flux.
We define $\tF^{(\bsf)}_V$ as the set of $\bk'\in\tF_V$ which 
is written as $\bk'=R_V(\bk)$ for some (not necessarily unique) $\bk\in\tF^{(\bsf)}$.

Note that an arbitrary $\bk\in\tF^{(\bsf)}$ is written as
\eq
\bk=(\bk_V,\bk_{\bV}),
\lb{kdec}
\en 
where $\bk_V=R_V(\bk)$ and $\bk_{\bV}=R_{\bV}(\bk)$.
We then have $\bk_V\in\tF^{(\bsf)}_V$, and $\bk_{\bV}\in\tF^{(-\bsf)}_{\bV}$.
We remark here that $\tF^{(-\bsf)}_{\bV}$ is the set of configurations on $\bV$ with incoming flux to $\bV$ (i.e., outgoing flux from $V$) equal to $-\bsf$. 

\subsection{$Z_N$ gauge theories}
We consider the $Z_N$ gauge theory,  generated by $\mathbb{Z}_N = \{ g^0=1, g^1, \cdots, g^{N-1} \}$,  where $g$ is a generator of the $Z_N$ and satisfies $g^N=1$ and $g^{-1}=g^\dagger$.

\paragraph*{Operators and states}
With each link $\eL$, we associate the $N$-dimensional Hilbert space $\calH_\ell$, whose orthonormal bra-basis is given by ${}_\ell\bra{U}$ with $U\in\mathbb{Z}_N$.
The coordinate operator $\hat U_\ell$ and the momentum (electric) operator $\hat E_\ell^g$ act on this bra-state as
\beqa
{}_\ell \langle U \vert \hat U_\ell &=&  {}_\ell \langle U \vert r_1(U), \quad 
{}_\ell \langle U \vert \hat E_\ell^g = {}_\ell\langle g U \vert, \quad
{}_\ell \langle U \vert \hat E_{\bar \ell}^g = {}_\ell \langle  U g^\dagger \vert,
\eeqa
where $r_1(U)$ is the fundamental representation of the $Z_N$ group such that $r_1(g_1g_2) =r_1(g_1) r_1(g_2)$ for $g_1,g_2\in \mathbb{Z}_N$. All irreducible representations are one dimensional and explicitly given by $r_k(g) = e^{i 2\pi k/N}$ for $ k=0,1,2,\cdots, N-1$.
  
The basic ket-state  $\vert h \rangle_\ell $ with $h\in \mathbb{Z}_N$ is defined as
\beqa
{}_\ell \langle U \vert h \rangle_\ell = \delta_{U,h}, 
\eeqa
and the general state can be expressed as
\beqa
\vert \Psi \rangle_\ell =\sum_{n=0}^{N-1} c_n \vert g^n \rangle_\ell, \quad  c_n \in \mathbb{C},
\eeqa
where $\vert g^n\rangle_l$ with $n\in\oN$ forms a basis of $\vert \Psi\rangle_\ell$.

We now introduce the basis of the flux representation as
\beqa
\vert k \rangle_\ell &=&\frac{1}{\sqrt{N}}\sum_{n=0}^{N-1} r_k(g^n)\vert g^n \rangle_\ell ,\qquad k\in\oN, 
\eeqa
which leads to
\beqa
{}_\ell \langle U \vert k \rangle_\ell &=& \frac{1}{\sqrt{N}}   \sum_{n=0}^{N-1}r_k(g^n) {}_\ell\langle U \vert g^n \rangle_\ell 
=\frac{1}{\sqrt{N}} r_k(U)\sum_{n=0}^{N-1} {}_\ell\langle U \vert g^n \rangle_\ell \nn \\
&=&  \frac{1}{\sqrt{N}}  r_k(U) , \qquad  \sum_{n=0}^{N-1} {}_\ell\langle U \vert g^n \rangle_\ell =1 .
\eeqa
Since
\beqa
\hat E_\ell^g \ket{k}_\ell &=& \frac{1}{\sqrt{N}}\sum_{n=0}^{N-1} r_k(g^n)\vert g^{n-1} \rangle_\ell
= r_k(g)  \ket{k}_\ell ,
\eeqa
so that $\ket{k}_\ell$ is an eigenstate of the electric operator $\hat E_\ell^g$ with an eigenvalue $r_k(g)$.
We shall use this electric flux representation, which is suited for studying reduced density matrices.

The Hilbert space $\calH$ for the whole system is spanned by the basis states
\eq
\ket{\bk}:=\bigotimes_{\eL}\ket{k_\ell}_\ell,
\lb{ketk}
\en
where $\bk\in\tF$.
The gauge invariant condition at $x$ that
\beqa
G_x^g \ket{\bk} = \ket{\bk}, \qquad
G_x^g:=\prodtwo{y\in\calS}{\mbox{ s.t. } (x,y)\in\calL}\hat E_{(x,y)}^g
\eeqa
leads to
\beqa
r_1(g)^{d_x^{\calL}(\bk)}=1 \Rightarrow d_x^{\calL}(\bk)=0 ,
\eeqa
where we use a property that $r_k(g) = r_1(g)^k$.
Therefore the divergence-free condition for $\bk$ corresponds to the gauge invariance of $\ket{\bk}$ at all $x\in\calS$.
In terms of link variables $U$, $\ket{\bk}$ represents $r_{k_\ell}(U_\ell)=r_1(U_\ell)^{k_\ell}$ at each link $\ell$, and the gauge invariant (divergence-free) condition means   that $\{ r_1(U_\ell)^{k_\ell}\}_{\eL}$ forms several closed loops with identifications  that $ r_1(U_\ell)^{N-k_\ell} = r_1(U_\ell^\dagger)^{k_\ell} $.

For a subset $V\subset\calL$ we also define $\calH_V$ as the space spanned by
\eq
\ket{\bk_V}_V:=\bigotimes_{\ell\in V}\ket{k_\ell}_\ell,
\en
where $\bk_V=(k_\ell)_{\ell\in V}\in\tF_V$.
The state $\ket{\bk_V}_V$ is not necessarily gauge invariant.

If $\bk_V\in\tF^{(\bsf)}_V$ and $\bk'_V\in\tF^{(\bsf')}_V$ with $\bsf\neq\bsf'$, the corresponding kets $\ket{\bk_V}_V$ and $\ket{\bk'_V}_V$ are orthogonal.
This means that the Hilbert space $\calH_V$ is decomposed into a direct sum
\eq
\calH_V=\bigoplus_{\bsf}\calH_V^{(\bsf)},
\lb{Hdec}
\en
where $\calH_V^{(\bsf)}$ is spanned by $\ket{\bk_V}_V$ with $\bk_V\in\tF^{(\bsf)}_V$.

Fix an arbitrary subset $V\subset\calL$ and let $\bV=\calL\backslash V$.
Corresponding to the decomposition \rlb{kdec} of $\bk\in\tF^{(\bsf)}$, the state \rlb{ketk} is decomposed as
\eq
\ket{\bk}=\ket{\bk_V}_V\otimes\ket{\bk_{\bV}}_{\bV},
\lb{ketkdec}
\en
where $\ket{\bk_V}_V\in\calH_V^{(\bsf)}$ and $\ket{\bk_{\bV}}_{\bV}\in\calH_{\bV}^{(-\bsf)}$.

\paragraph*{Reduced density matrix}
Take an arbitrary normalized state $\ket{\Psi}\in\calH$, and expand it as
\eq
\ket{\Psi}=\sum_{\bk\in\tF}\psi(\bk)\,\ket{\bk},
\lb{Psi1}
\en
where $\psi(\bk)\in\mathbb{C}$.
We shall fix a subset $V\subset\calL$ and its complement $\bV=\calL\backslash V$, and study the reduced density matrix in the region $V$ for the state $\ket{\Psi}$.

By taking into account the decomposition \rlb{Fdec} of $\tF$, and the decompositions \rlb{kdec}, \rlb{ketkdec} of $\bk$ and the corresponding ket, the state \rlb{Psi1} can be written as
\eqa
\ket{\Psi}&=\sum_{\bsf}\sum_{\bk\in\tF^{(\bsf)}}\psi(\bk)\,\ket{\bk}
\nl&
=\sum_{\bsf}
\sum_{\bk_V\in\tF^{(\bsf)}_V}\sum_{\bk_{\bV}\in\tF^{(-\bsf)}_{\bV}}
\psi(\bk_V,\bk_{\bV})\,\ket{\bk_V}_V\otimes\ket{\bk_{\bV}}_{\bV},
\lb{Psi2}
\ena
where the first sum is over admissible $\bsf$.
Then the corresponding density matrix is written as
\eq
\ket{\Psi}\bra{\Psi}=
\sumtwo{\bsf}{\bsf'}
\sumtwo{\bk_V\in\tF^{(\bsf)}_V}{\bk'_V\in\tF^{(\bsf')}_V}
\sumtwo{\bk_{\bV}\in\tF^{(-\bsf)}_{\bV}}
{\bk'_{\bV}\in\tF^{(-\bsf')}_{\bV}}
\psi(\bk_V,\bk_{\bV})\,\overline{\psi(\bk'_V,\bk'_{\bV})}
\,\,\ket{\bk_V}\bra{\bk'_V}\otimes
\ket{\bk_{\bV}}\bra{\bk'_{\bV}}.
\lb{density_matrix}
\en
Since $\ket{\bk_{\bV}}$ with $\bk_{\bV}\in\tF_{\bV}$ are orthonormal, the desired reduced density matrix is readily found to be
\eqa
\hrho_V&:={\rm Tr}_{\calH_{\bV}}\bigl[\ket{\Psi}\bra{\Psi}\bigr]
\nl
&=\sum_{\bsf}
\sumtwo{\bk_V\in\tF^{(\bsf)}_V}{\bk'_V\in\tF^{(\bsf)}_V}
\sum_{\bk_{\bV}\in\tF^{(-\bsf)}_{\bV}}
\psi(\bk_V,\bk_{\bV})\,\overline{\psi(\bk'_V,\bk_{\bV})}
\,\,\ket{\bk_V}\bra{\bk'_V}
\nl
&=\sum_{\bsf}p_{\bsf}\,\hrho_V^{(\bsf)}.
\lb{rhoV}
\ena
We have here defined the density matrix on $\calH^{(\bsf)}_V$ (see \rlb{Hdec}) by
\eq
\hrho_V^{(\bsf)}=\frac{1}{p_{\bsf}}
\sumtwo{\bk_V\in\tF^{(\bsf)}_V}{\bk'_V\in\tF^{(\bsf)}_V}
\sum_{\bk_{\bV}\in\tF^{(-\bsf)}_{\bV}}
\psi(\bk_V,\bk_{\bV})\,\overline{\psi(\bk'_V,\bk_{\bV})}
\,\,\ket{\bk_V}\bra{\bk'_V},
\lb{rhof}
\en
where $p_{\bsf}$ is obtained from the normalization condition.

As is well-known the final expression in \rlb{rhoV} implies
\eq
S[\hrho_V]=H[\boldsymbol{p}]+\sum_{\bsf}p_{\bsf}\,S[\hrho_V^{(\bsf)}],
\lb{SHS}
\en
where $H[\boldsymbol{p}]=-\sum_{\bsf}p_{\bsf}\log p_{\bsf}$ is the (classical) Shannon entropy for the probability distribution of the incoming flux through the boundary.
Note that the ``quantum part'' $S[\hat{\rho}_V^{\bsf}]$ is in general obtained by diagonalizing the expression \rlb{rhof}; this calculation may be nontrivial.

It may be suggestive to observe that, in the expression \rlb{SHS}, the von Neumann entropy $S[\hrho_V^{(\bsf)}]$ seems to reflect ``intrinsic entanglement'' between $V$ and $\bV$ while the Shannon entropy $H[\boldsymbol{p}]$ may simply reflect the behavior of Wilson loops that touch both $V$ and $\bV$.

\subsection{Some properties}
In the $Z_N$ gauge theories, the density matrix  can be expressed in the flux representation as
\eq
\rho = \sum_{\bk,\bk^\prime} \ket{\bk} \rho_{\bk,\bk^\prime} \bra{\bk^\prime}, 
\quad \ket{\bk}, \ket{\bk^\prime} \in {\cal H}_{\rm tot},
\en
in general\footnote{This form of the density matrix is more general than \rlb{density_matrix} for the pure state $\ket{\Psi}$.}, where ${\cal H}_{\rm tot}$ is the full Hilbert space without gauge invariance, and
$ \tr_{\rm tot}\, \rho=1$ implies $\sum_{\bk} \rho_{\bk\bk} = 1$.
The gauge invariance under  the gauge transformation $G_x^g$ and $G_y^{h}$ with $^\forall x, ^\forall y$
implies that $\rho_{\bk\bk^\prime} $ can be different from zero if and only if 
$d_x^{\cal L}(\bk) =d_y^{\cal L}(\bk^\prime)=0$ for $^\forall x, ^\forall y$.  This means that $\bk$ and $\bk^\prime$ are divergence-free.
Therefore
\eq
\tr_{\rm tot}\, \rho = \sum_{\bp}\sum_{\bk,\bk^\prime\in\tilde{\calF}} \langle \bp\vert \bk\rangle \rho_{\bk\bk^\prime}\langle \bk^\prime\vert \bp\rangle =\sum_{\bp\in \tilde{\calF}}\rho_{\bp\bp} =\tr_{\rm ph}\, \rho,
\en
where $\tr_{\rm ph}$ represents the trace over the physical space $\calH$.

Furthermore, the reduced density matrix is written as
\eq
\rho_V =\sum_{\bsf}\sumtwo{\bk_V\in\tF^{(\bsf)}}{\bk_V^\prime\in\tF^{(\bsf)}}\sum_{\bk_{\bV}\in\tF^{(-\bsf)}}
\rho(\bk_V,\bk_{\bV}; \bk_V^\prime, \bk_{\bV} ) \ket{\bk_V}\bra{\bk_V^\prime}.
\en
Therefore, for $\ket{\bp}_V\in H_{{\rm tot},V}$, we have
\eq
\rho_V \ket{\bp}_V = 0
\en
unless $\bp_V\in \tF_V$, so that
\eq
\tr_{{\rm tot}, V} \rho_V = \tr_V\, \rho_V =1,
\en
where $\tr_V$ is a trace over  ${\cal H}_V$ in \rlb{Hdec}.
In addition, we have
\eq
{}_{\bsf}\bra{\bk_V} \rho_V \ket{\bk_V^\prime}_{\bsf^\prime} =\delta_{\bsf,\bsf^\prime} \sum_{\bk_{\bV}\in\tF^{(-\bsf)}} \rho(\bk_V,\bk_{\bV}; \bk_V^\prime, \bk_{\bV} ). 
\en
Therefore we can extend $\rho_V$ in the full Hilbert space on V, $\calH_{{\rm tot},V}$ without any modifications.

The above argument shows that $\rho$ and $\rho_V$ can be regarded as the full and reduced density matrices in the full Hilbert spaces without gauge constraint.
The standard method then can be applied to prove  properties of $\rho_V$ such as positivity and strong sub-additativity\cite{ssa}. 

\subsection{Entanglement entropy for special states}
\paragraph*{Factorized states and the topological state}
Consider a special state in which the coefficients in \rlb{Psi1} and \rlb{Psi2} factorize as
\eq
\psi(\bk_V,\bk_{\bV})=\psi_V(\bk_V)\,\psi_{\bV}(\bk_{\bV})
\lb{factor}
\en
for any $\bk=(\bk_V,\bk_{\bV})\in\tF$.
Then the three summations in \rlb{rhof} can be treated independently to give
\eq
\hrho_V^{(\bsf)}=\frac{1}{p_{\bsf}}
\biggl(\sum_{\bk_{\bV}\in\tF^{(-\bsf)}_{\bV}}\hspace{-1.5mm}
|\psi_{\bV}(\bk_{\bV})|^2\biggr)
\biggl(\sum_{\bk_V\in\tF^{(\bsf)}_V}\hspace{-1mm}
\psi_V(\bk_V)\,\ket{\bk_V}\biggr)
\biggl(\sum_{\bk_V\in\tF^{(\bsf)}_V}\hspace{-1mm}
\overline{\psi_V(\bk_V)}\,\bra{\bk_V}\biggr),
\lb{rhof2}
\en
which shows that $\hrho_V^{(\bsf)}$ is pure, and hence $S[\hrho_V^{(\bsf)}]=0$.
In this case we find that the entanglement entropy $S[\hrho_V]$ is equal to the Shannon entropy $H[\boldsymbol{p}]$ for the probability distribution of the incoming flux $\bsf$.

The topological state, in which all the coefficients $\psi(\bk)$  in \rlb{Psi1} are identical, is an example where the factorization condition \rlb{factor} is satisfied. (See Refs.~\cite{wen,pre,CHR2013,Radicevic2014,Hamma2005,Hamma2005a} for related issues.)
This state is called the topological state, since an arbitrary (Wilson) loop has an unit eigenvalue.
Namely, for ${}^\forall\bk^\prime\in\tF$, we have
\beqa
\hat U^{\bk^\prime} \ket{{\rm topo}} = \ket {{\rm topo}} ,\quad
\hat U^{\bk^\prime} : = \prod_{\ell\in\Gamma}( \hat U_\ell)^{k^\prime_\ell},
\eeqa 
where 
\beqa
\ket{\rm topo} = \psi  \sum_{\bk\in\tF} \ket{\bk}
\eeqa
with $\psi$ is a complex number.
Indeed, since 
\beqa
{}_\ell\bra{U}\hat U_\ell \ket{k}_\ell = \frac{1}{\sqrt{N}} r_k(U)r_1(U) = 
 \frac{1}{\sqrt{N}} r_{k+1}(U)=
{}_\ell\langle{U} \ket{k+1}_\ell ,
\eeqa
where we use $r_{k_1}(U) r_{k_2}(U) = r_{k_1+k_2}(U)$,
we have
\beqa
\hat U^{\bk^\prime}   \ket{\rm topo} &=& \psi  \sum_{\bk\in\tF} \ket{\bk + \bk^\prime}
=  \psi  \sum_{\bk^{\prime\prime}\in\tF} \ket{\bk^{\prime\prime}}=\ket{\rm topo},
\eeqa
where $\bk^{\prime\prime} = \bk + \bk^\prime\in \tF$ . 

Writing $\alpha=|\psi|^2$, the expression \rlb{rhof2} becomes
\eq
\hrho_{{\rm topo},V}^{(\bsf)}=\frac{\alpha}{p_{\bsf}}
\biggl(\sum_{\bk_{\bV}\in\tF^{(-\bsf)}_{\bV}}1\biggr)
\biggl(\sum_{\bk_V\in\tF^{(\bsf)}_V}\ket{\bk_V}\biggr)
\biggl(\sum_{\bk_V\in\tF^{(\bsf)}_V}
\bra{\bk_V}\biggr),
\lb{rhof3}
\en
where
\beqa
p_{\bsf} &=& \alpha \biggl(\sum_{\bk_{\bV}\in\tF^{(-\bsf)}_{\bV}}1\biggr)
\biggl(\sum_{\bk_V\in\tF^{(\bsf)}_V} 1 \biggr) , \qquad
\sum_{\bsf} p_{\bsf} = 1 .
\lb{pf}
\eeqa
We shall argue that  $p_{\bsf}$ is independent of $\bsf$, and hence is equal to $1/N^{|\dV|-(n+m-1)}$,
where $N^{|\dV|-(n+m-1)}$ is a number of independent $\bsf$'s as shown in appendix \ref{app:dim_f}.
We thus get the desired result
\eq
S[\hat{\rho}_{{\rm topo},V}]=(|\dV|-n_\partial)\log N, \quad n_\partial:= n+m-1,
\en
where $|\dV|$ is a total number of boundary points, $n$ and $m$ are a number of  disconnected components of $V$ and $\bV$, respectively. 

The asserted independence is easily seen if one recalls the one-to-one correspondences between $\tF^{(\bsf)}$ with different $\bsf$.
Take admissible $\bsf_1$ and $\bsf_2$.
By restricting the map $\varphi_{1,2}$ to $\tF_V^{(\bsf_1)}$ and $\tF_{\bV}^{(-\bsf_1)}$, respectively, we obtain one-to-one correspondences between $\tF_V^{(\bsf_1)}$ and $\tF_V^{(\bsf_2)}$ and between $\tF_{\bV}^{(-\bsf_1)}$ and $\tF_{\bV}^{(-\bsf_2)}$.
We thus find that the expression \rlb{rhof3} for different $\bsf$ are in perfect one-to-one correspondences, so that a numbers of elements in both $\tF_V^{(\bsf)}$ and $\tF_{\bV}^{(-\bsf)}$ does not depend on $\bsf$, and thus  $p_{\bsf}$ is independent of $\bsf$ from \rlb{pf}.

\paragraph*{States with products of two loops}
We consider a simply entangled state, given by
\beqa
\langle U \vert \Psi \rangle &=& \frac{1}{\sqrt{n}}\sum_{i=1}^{n}  (U_{\Gamma_V})^{k_i}\,   (U_{\Gamma_{\bar V}})^{k_i}
\eeqa
for $ n \le N$, where integers $k_i$'s satisfy $k_i\not= k_j$ for $i\not=j$, and $\Gamma_V$ and $\Gamma_{\bV}$ are closed loops in $V$ and $\bV$ without touching the boundary, and $U_\Gamma$ is a product of $r_1(U)$ along the closed loop $\Gamma$. The reduced density matrix then becomes
\beqa
\rho^\Psi_V(U,U^\prime) &=& \frac{1}{n}\sum_{i=1}^n (U_{\Gamma_V})^{k_i}\,  (U^\prime_{\Gamma_V})^{k_i} ,
\eeqa
so that the entanglement entropy is given by
\beqa
S(V) &=& \log n .
\eeqa
In terms of the decomposition in eq.\rlb{SHS}, we have
\beqa
p_{\bsf} &=& \left\{
\begin{array}{ll}
1 & \mbox{ for } \bsf = {\bf 0} \\
0 & \mbox{ otherwise}  \\
\end{array}
\right. ,
\quad
\hat{\rho}_V^{(\bsf)} = \delta_{\bsf,{\bf 0}} \rho^\Psi_V ,
\eeqa
so that
\beqa
H[\bp]&=& 0,\qquad p_{\bsf} S[\hat{\rho}_V^{(\bsf)}] = \delta_{\bsf,{\bf 0}}\log n .
\eeqa

A simply disentangled state, on the other hand, is constructed as 
\beqa
\langle U \vert \Psi \rangle &=& \frac{1}{2}\left\{  U_{\Gamma_V}^{k_1}  +  U_{\Gamma_V}^{k_2} \right\}
\otimes
\left\{   U_{\Gamma_{\bar V}}^{k_1}  +    U_{\Gamma_{\bar V}}^{k_2} \ \right\},
\eeqa
which leads to
\beqa
\rho^\Psi_V(U, U^\prime) &=& \frac{1}{2} \left\{  U_{\Gamma_V}^{k_1}  +   U_{\Gamma_V}^{k_2} \right\}\
 \left\{ (U^\prime_{\Gamma_V})^{k_1}  + (U^\prime_{\Gamma_V})^{k_2} \right\}, \\
 S(V) &=& 0 .
 \eeqa

\paragraph*{Single-loop states}
An entangled loop state is constructed as
\beqa
\langle U \vert \Psi \rangle &=& \frac{1}{\sqrt{n}}\sum_{i=1}^n ( U_{\Gamma_V\Gamma_{\bar V}})^{k_i}
\eeqa
for $ n \le N$, where integers $k_i$'s satisfy $k_i\not= k_j$ for $i\not=j$, and
 $\Gamma_{V}\Gamma_{\bar V}$ is a closed loop  with $\Gamma_V\Gamma_{\bV}\in \bsf _0\not={\bf 0}$.
The reduced density matrix and entanglement entropy are given by
\beqa
\rho^\Psi_V(U, U^\prime) &=& \frac{1}{n}\sum_{i=1}^n   U_{\Gamma_V}^{k_i}\,  (U_{\Gamma_V}^\prime )^{k_i},\\
S(V) &=& \log n .
\eeqa
In terms of the decomposition in eq.\rlb{SHS}, we have
\beqa
p_{\bsf} &=& \left\{
\begin{array}{ll}
1/n & \mbox{ for } \bsf =\ {}^{\exists}k_i\bsf_0 \\
0 & \mbox{ otherwise }
\\
\end{array}
\right. ,
\quad
\hat{\rho}_V^{(k_i\bsf_0)} = U_{\Gamma_V}^{k_i}\,  (U_{\Gamma_V}^\prime )^{k_i}, 
\eeqa
so that
\beqa
H[\bp]&=& \log n,\qquad S[\hat{\rho}_V^{(\bsf)}] = 0 .
\eeqa

An example of a disentangled loop state is constructed as
\beqa
\langle U \vert \Psi \rangle &=& \frac{1}{\sqrt{2}}\left\{  (U_{\Gamma_V\Gamma_{\bar V}})^{k_1}  + 
 (U_{\Gamma^\prime_V\Gamma_{\bar V}})^{k_1}
 \right\},
\eeqa
which leads to
\beqa
\rho^\Psi_V(U, U^\prime) &=& \frac{1}{2} \left\{ (U_{\Gamma_V})^{k_1}  +   (U_{\Gamma^\prime_V})^{k_1} \right\}\
 \left\{  (U^\prime_{\Gamma_V})^{k_1} +  (U^\prime_{\Gamma^\prime_V})^{k_1} \right\}, \\
S(V) &=& 0 .
\eeqa

\subsection{One dimensional lattice without boundary}
\label{sec:one-dim}
Since one dimension is a little special, we here consider the $d=1$ case separately.

We consider ${Z}_N$-gauge theory on one dimensional lattice
with periodic boundary condition. Note that the open boundary is incompatible with the gauge invariance.
Since there are no Wilson loops (except one big loop on a whole lattice),
only the momentum operator $\hat{E}_\ell^g$
is a gauge invariant operator.

Considering the Gauss law, 
every link has the same electric eigenvalue.
Therefore, physical state is given by 
\begin{equation}
  |\Psi_k \rangle = \bigotimes_{l} | k\rangle_l
  =\ket{k}_V  \otimes \ket{k}_{\bar{V}} 
  \label{}
\end{equation}
with 
\begin{equation}
 \ket{k}_V \equiv \bigotimes_{l \in V} |k \rangle_l, \qquad
  \ket{k}_{\bar{V}}  \equiv \bigotimes_{l \in \bar{V}} |k \rangle_l,
  \label{}
\end{equation}
for arbitrary partitioning.
 
\paragraph*{Topological state}
A topological state is given by 
\begin{equation}
  |\mathrm{topo} \rangle = \frac{1}{\sqrt{N}}
  \sum_{k=0}^{N-1} | \Psi_k \rangle.
  \label{}
\end{equation}

The global density matrix becomes
\begin{eqnarray}
  \hat{\rho} &=&  |\mathrm{topo}\rangle
  \langle \mathrm{topo} | \nonumber \\
  &=& \frac{1}{N} \sum_{k,k'} | \Psi_k \rangle \langle \Psi_{k'}| \nonumber \\
  &=& \frac{1}{N} \sum_{k,k'} 
  \ket{k}_V \bra{k'}_V 
  \otimes  \ket{k}_{\bar{V}} \bra{k'}_{\bar{V}} 
  \label{}
\end{eqnarray}
and the reduced density matrix
\begin{equation}
  \hat{\rho}_V = 
  \mathrm{Tr}_{\bar{V}} \hat{\rho} 
  = \frac{1}{N} \sum_{k} \ket{k}_V  \bra{k}_V .
  \label{}
\end{equation}
Therefore, the entanglement entropy of the one dimensional
topological state is given by
\begin{eqnarray}
  S^\mathrm{(topo)}(V) &=&  - \mathrm{tr} \hat{\rho}_V \log \hat{\rho}_V \nonumber \\
  &=&  \log N \lb{ZN_1dim} \\
  &=& (n_B - n_\partial) \log N,
  \lb{eq:ee_1dim_topolo}
\end{eqnarray}
where  $n_B$  is the number of boundary points and $n_\partial = n+m-1$.
Since $n=m$ and $n_B=2n$, we always have 
\begin{equation}
  n_B - n_\partial = 1
  \label{}
\end{equation}
in one dimensional space.
The entanglement entropy does not depend on the number of links in $V$.
The result in \rlb{eq:ee_1dim_topolo} 
is the same as the topological state entropy formula in $d\ge 2$ lattice,

\paragraph*{General state}
We consider general state as
\begin{equation}
  |\alpha \rangle = 
  K \sum_{k=0}^{N-1} \alpha(k) | \Psi_k \rangle
  \label{}
\end{equation}
with the normalization coefficient
\begin{equation}
  K^2 = \frac{1}{\sum_{k=0}^{N-1} |\alpha(k)|^2}.
  \label{}
\end{equation}
The reduced density matrix is given by
\begin{equation}
  \hat{\rho}_{V} = K^2 \sum_{k} \alpha(k) \overline{\alpha(k)}  \ket{k}_V  \bra{k}_V.
  \label{}
\end{equation}
The entanglement entropy is given by
\begin{equation}
  S(V) = - \mathrm{tr} \hat{\rho}_V \log \hat{\rho}_V
  = - \sum p_k \log p_k
  \lb{one-dim}
\end{equation}
with $p_k \equiv |\alpha(k)|^2 / \sum|\alpha(k)|^2$.
For the topological state $p_0 = p_1 = \cdots = p_{N-1} = 1/N$,
\begin{equation}
  S^\mathrm{(topo)}(V) = \log N.
  \label{}
\end{equation}
For pure state $p_0 = 1, p_1 = \cdots = p_{N-1} = 0$,
\begin{equation}
  S^\mathrm{(pure)}(V) = 0.
  \label{}
\end{equation}

A simply entangled state
\begin{equation}
  |\Psi \rangle = \frac{1}{\sqrt{2}}
  \left\{ \ket{k}_V  \otimes \ket{k}_{\bar{V}} 
  + \ket{k'}_V  \otimes \ket{k'}_{\bar{V}}  \right\}
  \label{}
\end{equation}
with $k\not=k^\prime$, 
gives
\begin{equation}
  S(V) = \log 2 .
  \label{}
\end{equation}

\subsection{Relation to other proposals}
We here discuss relations of our definition of entanglement entropy  (or the reduce density matrix) for gauge theories, in particular, the $Z_N$ gauge theory to other proposals.

Our definition is equivalent to the electric boundary condition(electric center) in Ref.~\cite{CHR2013} and in Ref.~\cite{Radicevic2014},   to the extension of the Hilbert space in Ref.~\cite{BP2008}, and to the extended lattice construction in Ref.~\cite{Donnelly2012}. In this definition, the reduce density matrix $\rho_V$, from the whole density matrix $\rho$ restricted to the region $V$, satisfies
\beqa
\langle {\cal O}_V \rangle := \tr [ {\cal O}_V \rho ] = \tr_V  [ {\cal O}_V \rho_V ] 
\eeqa
for ${}^\forall {\cal O}_V \in {\cal A}_V$, where ${\cal A}_V$ is the set of gauge invariant operators on $V$, generated by $\hat E_\ell^g$ with $\ell\in V$ and $\hat U_p$ with the plaquette whose links are all included in $ V$, and $\tr_V$ is the trace over $\mathcal{H}_V$. It is noted that ${\cal A}_V$ is the maximal gauge invariant algebra on $V$.

\bigskip

The trivial center definition in Ref.~\cite{CHR2013}, denoted by $\rho_V^0$, is equivalent to the gauge fixed theory where the boundary links in the maximal tree are all fixed to the unit element. In this case, however, the set of gauge invariant operators ${\cal A}_V^0$, generated by $\hat E_\ell^g$ with $\ell\in V \backslash\{ \mbox{maximal tree} \}$ and the same set of plaquette $\hat U_p$ on $V$, 
is smaller than ${\cal A}_V$.  Similarly, the algebra $A_V^m$ associated with the magnetic center\cite{CHR2013,Radicevic2014}
 is smaller than $A_V^0$ .
 Therefore both $A_V^0$ and $A_V^m$ do not represent the region $V$ algebraically, 
so that  definitions based on the trivial center and the magnetic center are inadequate for  the entanglement entropy or the reduced density matrix on the region $V$.

\bigskip 

 In conclusion, our definition of the entanglement entropy or  reduced density matrix gives the unique definition of these quantities on the region $V$, in the sense that our reduced density matrix is associated with the maximally gauge invariant algebra $A_V$ on $V$.

\subsection{Gauge fixing}
Since the reduced density matrix $\rho_V$ does not  have the full gauge invariance as mentioned before,
the entanglement entropy may depend on whether gauge fixing is employed or not in the calculation, and on the choice of the gauge if the gauge fixing is used. 
In this subsection,  using a simple example, we explicitly demonstrate that the entanglement entropy with some gauge fixing is different from the one calculated without gauge fixing.

We consider the $Z_N$ gauge theories in one dimension with periodic boundary condition in subsection~\ref{sec:one-dim}.
Without gauge fixing, the entanglement entropy  is given in \rlb{one-dim} as
\be
S(V)= -\sum_k p_k\log p_k,  \qquad p_k=K^2\vert\alpha(k)\vert^2,\quad K^2=\frac{1}{\sum_k \vert\alpha(k)\vert^2}
\lb{eq:one-dim2}
\ee
for a general state
\be
\ket{\alpha} =K \sum_k \alpha(k) \ket{\Psi_k} .
\ee

Take $L$ lattice points on the circle as $\calS=\{1,2,\cdots, L\}$ and $\calL=\{(1,2),(2,3),\cdots, (L,1) \}$.
Links in the region $V$ are given by $\calL_V=\{(1,2),(2,3),\cdots, (L_V-1,L_V) \}$, while
those in $\bV$ by $\calL_{\bV}=\{(L_V,L_V+1),\cdots, (L,1) \}$, where $0 < L_V < L$ and  $\dV =\{ 1, L_V\}$.
Using gauge transformations on all points in $\calS$ except one, we can always make $U_\ell = 1$ for all $\ell\in \calL$ except one $\ell$ which may be in $\calL_V$ or $\calL_{\bV}$. In any cases, the reduced density matrix from the global pure state is always pure, so that the entanglement entropy is always zero.
This is clearly different from \rlb{eq:one-dim2} without gauge fixing.

We next consider the gauge fixing using all points in $\calS$ except $\dV$.
In this case we can make $U_\ell = 1$ for all $\ell\in \calL$ except two $\ell$'s, one $\ell$ in $\calL_V$ and the other in  $\calL_{\bV}$. For example, we can take $U_{(1,2)}\not= 1$ and $U_{(L,1)}\not=1$.
Since the gauge invariance still holds on the site $1$, the physical state can be written as
\be
\ket{\alpha} = K \sum_k \alpha(k) \ket{k}_{(1,2)}\otimes \ket{k}_{(L,1)} .
\ee
Then the reduce density matrix is given by
\be
\hat\rho_V = K^2\sum_k \vert\alpha(k)\vert^2  \ket{k}_{(12)}\ {}_{(12)}\bra{k},
\ee
which leads to \rlb{eq:one-dim2}  for the entanglement entropy. For the topological state,
it reducers to
\beqa
S^{\rm (topo)}(V) = \log N .
\eeqa

The above consideration leads to an important lesson that the entanglement entropy does not depend on the gauge fixing if and only if points in $\dV$ are excluded in the gauge fixing (including no gauge fixing at all).  Otherwise, the entanglement entropy does depend on the gauge choice.

\section{Conclusion}
We have proposed the definition of the entanglement entropy in lattice gauge theories for an arbitrary subset of links 
not only in abelian theories but also in non-abelian theories,  
and explicitly given the replica formula based on our definition.
In the $Z_N$ gauge theories, 
we  have expressed the whole Hilbert space by the flux representation basis states which are eigenstates of 
the gauge transformations. 
By using these basis states, we have explicitly argued that all the standard properties of entanglement entropy hold in our definition and calculated the entanglement entropy for topological states as
\be
S[\hat \rho_{\rm topo}, V] =(\vert \dV\vert -n_\partial ) \log N .
\ee
We have also found that the entanglement entropy depends on the gauge fixing in general.

It will be important to extend our analysis for the $Z_N$ gauge theories to non-abelian gauge theories, since
our definition is applicable also  to non-abelian cases without any difficulties.
In order to calculate  the entanglement entropy analytically in non-abelian gauge theories, 
we need some useful basis such as the flux representation in the $Z_N$ gauge theories.  
In the $Z_N$ gauge theories, 
the flux representation basis diagonalizes gauge transformations simultaneously. 
On the other hand, in non-abelian gauge theories, 
gauge transformations cannot be diagonalized simultaneously 
since they do not commute each other. 
We therefore need some new ideas for non-abelian gauge theories. 
In appendix~\ref{app:general}, some analyses in this direction are given. For example,
the entanglement entropy for the topological state in one dimension is calculated as
\be
S_V = \log \vert G \vert 
\ee 
in the discrete non-abelian gauge theories, where $\vert G\vert$ is a number of elements of the discrete group.

Others directions in future investigations  include  perturbative calculations for the entanglement entropy in gauge theories\cite{P1, P2, P3,Huang:2014pfa} without gauge fixing at boundaries and numerical simulations for the entanglement entropy in lattice gauge theories\cite{BP2008, Nakagawa:2009jk,Nakagawa2011}.

\vskip 1cm

After completing our investigations presented in this report, we noticed a paper\cite{Ghosh2015} in which the authors also propose the definition of the entanglement entropy in lattice gauge theories.  We find that their proposal is identical to ours, though research directions in this paper are somewhat different from theirs.
See also Ref.~\cite{Hung:2015fla} for a related result.

\section*{Acknowledgement}
The authors would like to thank the Yukawa Institute for Theoretical Physics at Kyoto University, where this work was initiated during the YITP-W-14-08 on ``YITP Workshop on Quantum Information Physics (YQIP2014)".
We also thank Dr.~K.~Kikuchi for discussions, and M. N. thanks Dorde Radi$\check{c}$evi$\acute{c}$ for stimulating discussions.
This work is supported in part by  the JSPS Grant-in-Aid for Scientific Research (Nos. 25287046, 25287046, 25400407) , the MEXT Strategic Program for Innovative Research (SPIRE) Field 5, and Joint Institute for Computational Fundamental Science (JICFuS).
M. N. and T. N. are supported by the JSPS fellowship.

\appendix
\section{The number of admissible $\bsf$}
\label{app:dim_f}
Here  the whole $\calL$ is assumed to be finite and connected.

Suppose that $V$ and $\bV$ are decomposed into connected components as
\eq
V=V_1\cup\cdots\cup V_n,\quad \bV=\bV_1\cup\cdots\cup \bV_m,
\en
with $n,m\ge1$.
Consequently the boundary $\dV=\dbV$ is decomposed as
\eq
\dV=\dV_1\cup\cdots\cup\dV_n,\quad \dbV=\dbV_1\cup\cdots\cup\dbV_m,
\en
where $\dV_i$ and $\dbV_j$ are the boundaries of $V_i$ and $\bV_j$, respectively; they may not be connected.

Let us denote by 
\eq
\calF:=\set{\bsf=(f_x)_{x\in\dV}}{$f_x\in\oN$ for $x\in\dV$}
\en
the set of all configurations of incoming currents (including ``unphysical'' ones).
We have $|\calF|=N^{|\dV|}$.

The necessary and sufficient conditions for the admissibility of $\bsf$ are
\eq
\sum_{x\in\dV_i}f_x=0,\quad\text{and}\quad
\sum_{x\in\dbV_j}f_x=0,
\lb{cond}
\en
for all $i=1,\ldots,n$ and $j=1,\ldots,m$.
There are $n+m$ constraints, but they are not independent.
To see this note that any $\bsf$ satisfies
\eq
\sum_{i=1}^n\sum_{x\in\dV_i}f_x=\sum_{j=1}^m\sum_{x\in\dbV_j}f_x,
\lb{ff}
\en
because $\dV=\dbV$.
We thus see that
\eq
\sum_{x\in\dbV_m}f_x
=\sum_{i=1}^n\sum_{x\in\dV_i}f_x-\sum_{j=1}^{m-1}\sum_{x\in\dbV_j}f_x
\lb{constraint_full}
\en
holds for any $\bsf$.
It is therefore sufficient consider the constraints \rlb{cond} for $i=1,\ldots,n$ and $j=1,\ldots,m-1$.
There are $\gamma:=n+m-1$ constraints.

To count a number of admissible $\bsf$, 
let us introduce matter fields (or external sources) with $Z_N$ charge
on lattice sites $\{ x \}_{x\in\calS}$.
For a given charge density distribution $\left\{ q(x) \right\}_{x\in\calS}$,
an admissible flux in this general case is determined so as to satisfy the Gauss law as
\begin{equation}
  \sum_{x \in \partial V_{i}} f_x  = Q_i \in\oN,
 \lb{constraint_1} 
\end{equation}
for each $i = 1, \dots, n$, and
\begin{equation}
  \sum_{x \in \partial \bar{V}_{j}} f_x   = - Q_{j+n} \in\oN, 
 \lb{constraint_2}   
\end{equation}
for each $j = 1, \dots, m$, 
where $Q_k$ ($k=1,\cdots, \gamma+1$), a sum of $q(x)$ over inner point, is 
 a total charge inside the region $V_i$ or $\bar{V}_j$ excluding  boundaries.
The minus sign in the second equation comes from the fact that a flux on $\dbV$
has a relative minus sign  with respect to a flux on $\dV$. 
Due to the constraint \rlb{constraint_full},  we have
\beqa
Q_{\gamma+1}=-\sum_{k=1}^\gamma Q_k,
 \lb{constraint_3} 
\eeqa
so that only $Q_1,\cdots, Q_\gamma$ are independent. 
We then define $\calF_{Q_1,\ldots, Q_\gamma}$ as the set of $\bsf\in \calF$  which satisfies
\rlb{constraint_1} and \rlb{constraint_2}. 
It is then easy to see
\eq
\calF=\bigcup_{Q_1,\ldots,Q_\gamma\in\oN}
\calF_{Q_1,\ldots, Q_\gamma}.
\lb{Fdec}
\en
Note that $\calF_{0,\ldots,0}$ is the set of admissible $\bsf$'s that we are interested in.

\begin{figure}
\centerline{\epsfig{file=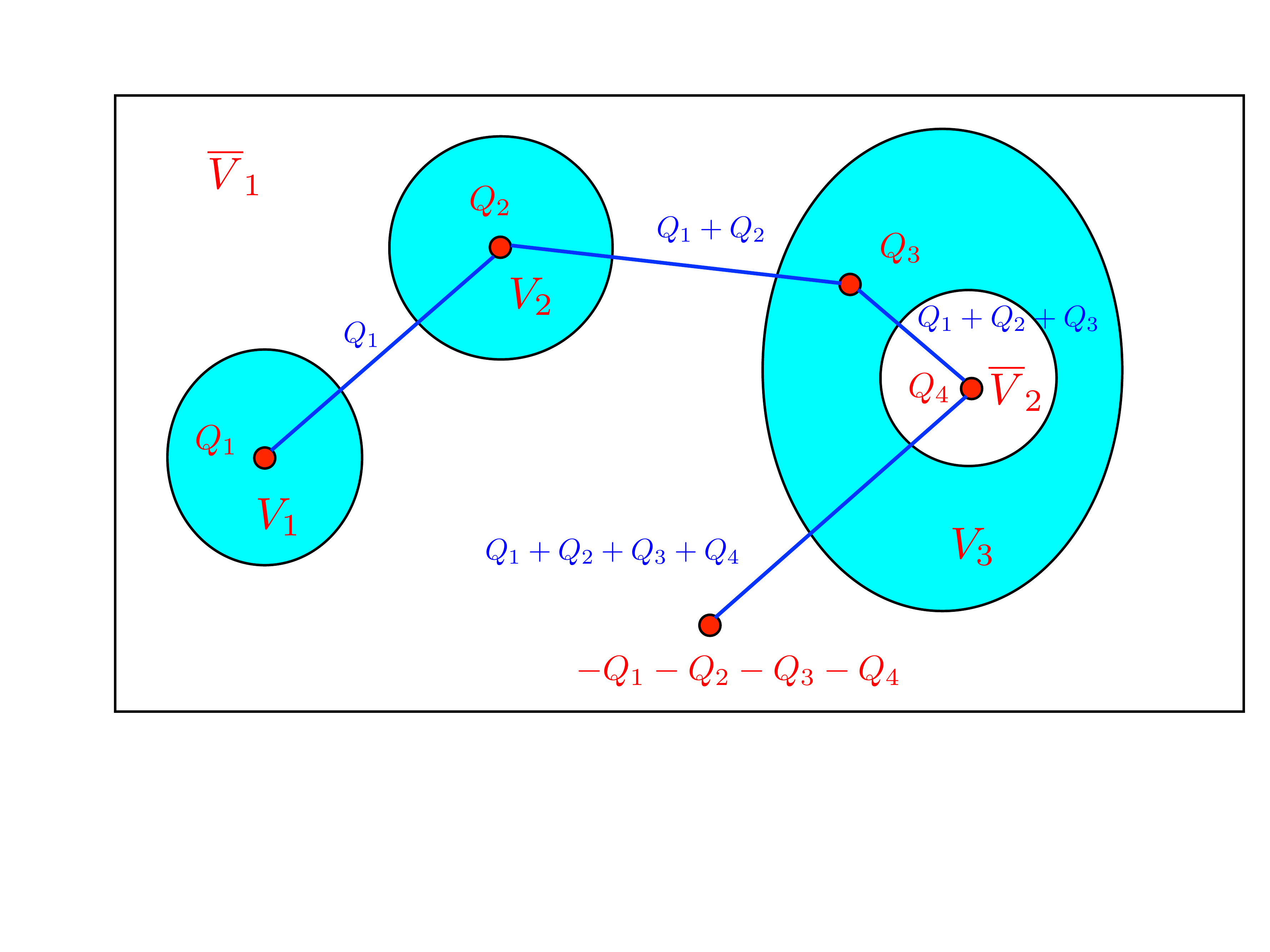,width=10cm}}
\caption{
An example of the connection and charge distributions.
}
\label{f:connection}
\end{figure}

Now we will argue that
$\mathcal{F}_{0,\cdots,0}$ is isomorphic to $\mathcal{F}_{Q_1, \cdots, Q_{\gamma}}$ for an arbitrary $Q_1,\cdots,Q_\gamma$.
Take one internal point $x_k$ from each region $V_i$ or $\bV_j$.
Connect these points by the following condition: (1) links can be used once. (2) except start and end points, each point belongs to only two links (3) the end point is always $x_{\gamma+1}$. 
It is easy to see such a connection always exist. 
By changing the order of point  $x_k$ along this connection and renaming $x_k$ in this order, we write  the connection as $\Gamma_{x_1x_2}\Gamma_{x_2x_3} \cdots \Gamma_{x_{\gamma-1} x_\gamma}\Gamma_{x_\gamma x_{\gamma+1}}$, where $\Gamma_{x_i x_i+1}$ is a set of links which connect $x_i$ and $x_{i+1}$. 
For an illustration, see Fig.~\ref{f:connection}.

For ${}^\forall \{ Q_1, \cdots, Q_\gamma \}$ (this is also reordered),
 we  define $\bk^{ Q_1, \cdots, Q_\gamma }$ on a link $\ell$ as
\beqa
k^{ Q_1, \cdots, Q_\gamma }_\ell &=&\left\{ 
\begin{array}{ll}
\sum_{i=1}^{k} Q_i, & \mbox{ for } \ell \in \Gamma_{x_kx_{k+1}} \\
0, & \mbox{otherwise} \\
\end{array}
\right. .
\eeqa
See Fig.~\ref{f:connection} again as an example. A blue letter such as $Q_1+Q_2$ represents a charge  on some lines, while
a red letter such as $Q_k$ is a charge on the point $x_k$.
Note that the net charge flowing  out from the $k$-th region (some $V_i$ or $\bV_j$ ) is equal to
$\sum_{i=1}^{k} Q_i -\sum_{i=1}^{k-1} Q_i = Q_k$. 
It is then easy to see that the map for $\bk \in \mathcal{F}_{0,\cdots,0}$ defined by
\beqa
\varphi^{ Q_1, \cdots, Q_\gamma }_{0,\cdots,0}(\bk) : = \bk + \bk^{ Q_1, \cdots, Q_\gamma }
\eeqa
establishes an isomorphism from   $\mathcal{F}_{0,\cdots,0}$
to $\mathcal{F}_{Q_1, \cdots, Q_{\gamma}}$.
This proves the number of $\mathcal{F}_{Q_1, \cdots, Q_{\gamma}}$ is independent 
of $Q_1, \cdots, Q_\gamma$.

A number of possible charge distribution $\left\{ Q_1, Q_2, \cdots, Q_\gamma \right\}$ 
is $N^\gamma = N^{n+m-1}$.  Therefore, for any charge distributions $\{q(x)\}_{x\in\calS}$ including $\{q(x)\}_{x\in\calS} =\{0\}_{x\in\calS}$,  
the total number of the admissible $\bsf$ is $N^{|\partial V| - (n+m-1)}$.

\section{Entanglement entropy for non-abelian gauge theories}
\label{app:general}
\subsection{About the Hilbert space on a link}

We generalize the formulation of the $Z_n$  case to non-abelian gauge theories. 
We take a group $G$ which we assume to be a compact group. We define the momentum operator $L_\ell(g )$ and the position operator $U^{\pi}_\ell $ via 

\be
\bra{U}L_\ell(g )\ket{\Psi} = \Psi(g ^{-1}U) , \ \ \ \bra{U} (U^{\pi}_\ell)^{\alpha}{}_{\beta}\ket{\Psi} = \pi (U)^{\alpha} {}_{\beta} \Psi(U)
\ee
where $g \in G$ and $\pi $ is a representation of $G$. If we inverse the direction of link $\ell$, the operator $L_{\ell^T}(g)$ and $U_{\ell^T}^{\pi}$ is defined as follows:
\be
\bra{U}L_{\ell^T}(g)\ket{\Psi} = \Psi(U g) , \ \ \ \bra{U} (U_{\ell^T} ^{\pi})^{\alpha}{}_{\beta} \ket{\Psi} = \pi(U^{-1})^{\alpha}{}_{\beta}\Psi(U) = (\pi(U)^*)_{\beta} {}^{\alpha} \Psi(U)
\ee

It is known that the $L^2 $ space on a group $G $ (square integrable functions over $G$) decomposes to the direct sum of $\pi^ {\dagger} \boxtimes \pi$ which is a irreducible representation of $G \times G$ as follows\cite{CMS1995}:
\be
L^2 (G ) \simeq \bigoplus _{\pi \in \text{Irr}(G)} V_{\pi} ^{\dagger} \otimes V_{\pi} ,
\lb{decomposition}
\ee
where we denote $\pi$ as an (unitary) irreducible representation of $G$ and $\text{Irr}(G)$ as the set of irreducible representation and $\pi ^{\dagger}(g ) = {}^t \pi (g ^{-1})$ is the dual representation. 
The meaning of \rlb{decomposition} will become clear below.

We first consider the basic state $\ket{\pi^\alpha{}_\beta}$ defined via
\be
\braket{U|\pi^{\alpha}{}_{\beta}} = \pi^{\dagger}(U)_{\beta} {}^{\alpha},
\ee
with which we can explicitly write the action of $L_\ell(g)$ as
\ba
\bra{U} L_\ell(g) \ket{\pi ^{\alpha}{}_{\beta}} &=& \pi^{\dagger}(g^{-1}U)_{\beta}{}^{\alpha}= \pi(U^{-1} g) ^{\alpha }{}_{\beta} 
= \pi(g)^{\gamma} {}_{\beta} \braket{U | \pi^{\alpha }{}_{\gamma}}.
\ea
Therefore we have
\ba
L_\ell(g)\ket{\pi ^{\alpha}{}_{\beta}} &=& \ket{\pi^{\alpha }{}_{\gamma}}  \pi(g)^{\gamma} {}_{\beta}, \\
L_{\ell^T}(g)\ket{\pi ^{\alpha}{}_{\beta}} &=& \pi(g^{-1})^{\alpha} {}_{\gamma} \ket{\pi^{\gamma }{}_{\beta}} . 
\ea

The dual vector of $\ket{\pi ^{\alpha}{}_{\beta}}$is given by $(\ket{\pi^{\alpha}{}_{\beta}})^{\dagger} = \bra{\pi ^{\beta}{}_{\alpha}}$, and the projection operator to the 
subspace $\pi ^{\dagger} \boxtimes \pi$ is given by 
\be
P_{\pi} = \dim V_{\pi} \sum_{\alpha, \beta} \ket{\pi^{\alpha}{}_{\beta} }\bra{\pi ^{\beta} {}_{\alpha}}.
\lb{projection}
\ee
The factor $\dim V_{\pi}$ is needed here since the normalization of vector$\ket{\pi ^{\alpha}{}_{\beta}}$ is given by
\be
\braket{\pi^{\alpha}{}_{\beta}|\pi^{\gamma}{}_{\delta}} = \int d U \ \pi^{\alpha}{}_{\beta}(U) \pi^{\gamma}{}_{\delta}(U^{-1}) = \frac{\delta^{\alpha} {}_{\delta} \delta^{\gamma}{}_{\beta}}{\dim V_{\pi}} 
\ee
By the projection \rlb{projection}, the meaning of \rlb{decomposition} becomes clear.

\subsection{Gauge invariant states}
In the lattice gauge theory, the total Hilbert space $\mathcal{H}_0$ is $\bigotimes _l (L^{2}(G))_l$.  The physical Hilbert space $\mathcal{H}$ as the subspace of $\mathcal{H}_0$ is consist of gauge invariant states,
which satisfy
\be
G_x^g \ket{\Psi} = \ket{\Psi}, \qquad G_x^g:=\prodtwo{y\in\calS}{\mbox{s.t. } \ell=(x,y)\in\calL}L_{\ell }(g)  ,
\ee
at  ${}^\forall x\in\calS$.
The basis state in $\calH_0$ is written in general as
\ba
\ket{\boldsymbol{\pi}^{\boldsymbol{\alpha}}{}_{\boldsymbol{\beta}}}&:=& \bigotimes_{\ell \in \calL}\ket{(\pi_\ell)^{\alpha_\ell}{}_{\beta_\ell}}
\lb{basic_state}
\ea
where $\pi_\ell$ indicates an irreducible representation of $G$ on a link $\ell$.

Unlike the $Z_N$ gauge theories, it is not so easy to write gauge invariant conditions for the state in \rlb{basic_state}.   Let us consider the one dimensional case as a simplest example.
In this case, the nontrivial part of the gauge invariant condition at $x$ becomes 
\be
\ket{\pi^{\alpha_1}{}_{\gamma_1}}_{\ell_1}\ \pi (g)^{\gamma_1}{}_{\beta_1}\otimes  \pi^\prime(g^{-1})^{\alpha_2}{}_{\gamma_2}\  \ket{\pi^\prime\, {}^{\gamma_2}{}_{\beta_2}}_{\ell_2} 
= \ket{\pi^{\alpha_1}{}_{\beta_1}}_{\ell_1}\otimes \ket{\pi^\prime\,{}^{\alpha_2}{}_{\beta_2}}_{\ell_2}
\ee
where $\ell_1=(x-1,x)=(x,x-1)^T$ and $\ell_2=(x,x+1)$. Integrating this equation over $g$ with $\int dg = 1$,
we find that a gauge invariant state at $x$ has a form as 
\be
\frac{1}{\dim V_\pi}  \ket{\pi^{\alpha}{}_{\gamma}}_{\ell_1} \otimes \ket{\pi^{\gamma}{}_{\beta}}_{\ell_2} ,
\ee
where two irreducible representations on $\ell_1$ and $\ell_2$ must be equal.

In higher dimensions, however, the condition becomes more complicated.
On a $d$-dimensional  hyper-cubic lattice, the gauge invariant condition at $x$ reads
\ba
\int dg \prod_{\mu=1}^d \ket{\pi_{\mu}{}^{\alpha_\mu}{}_{\gamma_\mu}}_{\ell_\mu}\ \pi_\mu (g)^{\gamma_\mu}{}_{\beta_\mu}\otimes  \pi_{\bar\mu}(g^{-1})^{\alpha_{\bar\mu}}{}_{\gamma_{\bar\mu}}\  \ket{\pi_{\bar\mu}\, {}^{\gamma_{\bar\mu}}{}_{\beta_{\bar\mu}}}_{\ell_{\bar\mu}} &=&
 \prod_{\mu}  \ket{\pi_\mu{}^{\alpha_\mu}{}_{\beta_\mu}}_{\ell_\mu}\otimes \ket{\pi_{\bar\mu}\,{}^{\alpha_{\bar\mu}}{}_{\beta_{\bar\mu}}}_{\ell_{\bar\mu}} ,\notag \\
 \lb{general_condition}
\ea
where $\ell_\mu = (x,x+\mu)$ and $\ell_{\bar\mu} = (x,x-\mu)$.
This implies that a product of $2d$ irreducible representations of $\pi_\mu$ and $\pi_{\bar\mu}$ must contain the trivial representation.  For example, in the case of SU(2) gauge group at $d=2$, 4 non-negative integers $k_{1,2,3,4}$,
which are numbers of boxes in the SU(2) Young tableaux and specify  irreducible representations of SU(2),
must satisfy
\beqa
\{\vert k_1-k_2\vert, 
\cdots, k_1+k_2 \} \cap \{\vert k_3-k_4\vert, 
\cdots, k_3+k_4 \} \not= \emptyset . \notag
\eeqa
For general gauge groups in higher dimension, it is hard to find a simple condition for \rlb{general_condition}.

\subsection{Examples}
As was seen in the previous subsection, it is not so easy to construct general gauge invariant states in higher dimensions. Therefore, in this subsection, we  consider two examples at $d=1$.

\subsubsection{One dimensional topological state with periodic boundary condition}

Assume that there are $N$ links on a circle({\it i.e.} the periodic boundary condition). In this boundary condition, similar results are obtained by Donnelly\cite{Don} in the theories defined on the continuum space. The physical Hilbert space is given by the gauge invariant functions. From the analysis in the previous subsection, the basis are given by the characters of irreducible representations as
\ba
\ket{\psi} = \sum_{\pi \in \text{Irr }(G)} \psi(\pi)\ket{\pi} , \ \ \ \ket{\psi} \in  \mathcal{H} \label{physical} \\
\ket{\pi} =  \ket{\pi ^{\alpha_1} {}_{\alpha _2}} \otimes  \ket{\pi ^{\alpha_2} {}_{\alpha _3}} \otimes \cdots \otimes \ket{\pi^ {\alpha_{N}} {}_{\alpha _1}} , \qquad \braket{\pi\vert \pi} = 1.
\ea
The value of $\ket{\pi}$ at $\ket{U_1 ,\cdots, U_N} = \ket {U_1} \otimes \cdots \otimes \ket{U_N} $ becomes as follows.
\ba
\braket{U_1 , \cdots , U_N | \pi} &=& \pi ^{\dagger}(U_1)_{\alpha_2}{} ^{\alpha_1} \pi ^{\dagger}(U_2)_{\alpha_3 }{}^{\alpha_2} \cdots \pi^{\dagger}(U_N)_{\alpha_1} {}^{\alpha_N} \notag \\
&=& \text{\tr} ( \pi^{\dagger}(U_N \cdots U_1)) .
\ea 
As we have done in the abelian cases, to divide the physical Hilbert space into the tensor product of Hilbert spaces on the region $V$ and $\bar{V}$, we embed the physical Hilbert space $\mathcal{H}$ into a larger Hilbert space $\mathcal{H}'$ where \footnote{Unlike the $Z_N$ gauge theories in the main text, we here consider the minimum extension where gauge invariance is abandoned only at boundaries.} 
\ba
\mathcal{H}'  &=& \bigoplus_{\pi, \pi'}\bigoplus_{\bm{\alpha} , \bm{\beta} , \bm{\alpha}' \bm{\beta} '} \mathcal{H}_V^{(\pi) \bm{\alpha}}{}_{\bm{\beta}} \otimes \mathcal{H}_{\bar{V}}^{(\pi') \bm{\beta} '}{}_{\bm{\alpha}'} .
\ea
Here $\bm{\alpha} = (\alpha _1 , \cdots, \alpha _M) , \bm{\beta} = (\beta_1 , \cdots , \beta _M) $ are labels of boundaries when the subsystem $V$ is consist of $M$ intervals and 
$\bm{\alpha '} ,\bm{\beta '}$ are the corresponding ones in $\bV$.
Then we trace over $\mathcal{H}_{\bar{V}}^{({\pi}) \bm{\beta}}{}_{\bm{\alpha}}$,
regarding  the physical wave function $\ket{\psi}$ as an element of $\mathcal{H}'$.

As the simplest case, we consider $V$ (and $\bar{V}$) is an interval. In this case, the basic is written as 
\be
\ket{\pi} = (\dim V_{\pi}) ^{-1} \times \sum_{\alpha, \beta} \sqrt{\dim V_{\pi}} \ket{\pi_{V} ^{\alpha}{}_{\beta}} \otimes \sqrt{\dim V_{\pi}} \ket{\pi_{\bar{V}}^{\beta}{}_{\alpha}}.
\ee 
The reduced density matrix for physical wave function (\ref{physical}) is given by
\be
\rho_{V} = \sum _{\pi \in \text{Irr}(G)} p(\pi) (\dim V_{\pi}) ^{-2} \sum_{\alpha , \beta} (\sqrt{\dim V_{\pi}}\ket{\pi_V ^{\alpha}{}_{\beta}} )( \sqrt{\dim V_{\pi}}\bra{\pi_V^{\beta} {}_{\alpha}}) \label{rdm}
\ee
where $p(\pi) = |\psi(\pi) |^2$.  Its entanglement entropy is given by 
\ba
S_V &=& -\sum_{\pi \in \text{Irr}(G), \alpha, \beta}  p(\pi)(\dim V_{\pi} )^{-2} \log (p(\pi)(\dim V_{\pi} )^{-2}) \notag \\
&=& \sum_{\pi \in \text{Irr}(G)}p(\pi)(- \log p(\pi) + 2 \log \dim V_{\pi} ) \label{eenonabg}
\ea

Using the above result, we compute an entanglement entropy of  the topological state in finite non-abelian group $G$.
The topological state is given by 
\be
\ket{\text{topo}} = \frac{1}{\sqrt{|G|^{2N + 1}}}\sum_{g _i \in G} \ket{  g _1^{-1} g_N   } \otimes \ket{g _2 ^{-1} g _1  } \otimes \cdots \otimes \ket{g _{N} ^{-1} g _{N-1} },
\ee
where $\vert G\vert$ is the number of the element of $G$, and states satisfy $\braket{g\vert h} = \vert G\vert \delta_{g,h}$.  Here
$\ket{\text{topo}}$ is written as the element of $\mathcal{H}_0$, though it is gauge invariant, and
the coefficient $\psi(\pi)$ is  given by
\be
\psi (\pi) = \braket{\pi | \text{topo}} = \frac{\dim V_{\pi}}{ \sqrt{|G|}} ,
\ee
which leads to $ p(\pi) = (\dim V_{\pi}) ^2 /|G|$. Thus the entanglement entropy is calculated as
\be
S_V = \sum_{\pi \in \text{Irr}(G)} \frac{(\dim V_{\pi})^2}{|G|} \Big( - \log \frac{(\dim V_{\pi})^2 }{|G|} + 2 \log \dim V_{\pi} \Big) = \log |G| ,
\ee
where we use the identity $\sum_{\pi \in \text{Irr}(G)} (\dim V_{\pi})^2  = |G| $.
This result agrees with \rlb{ZN_1dim} for  the $Z_N$ gauge theories

\subsubsection{One dimensional topological state with open boundary condition}
Next we consider the case with  open boundary condition. From the gauge invariance of the bulk,  physical wave functions are given by a linear combination of functions on $G$ as
\ba
\ket{\psi} &=& \sum_{\pi \in \text{Irr}(G), \alpha, \beta  } \psi(\pi)_{\alpha}{}^{\beta} \sqrt{\dim V_{\pi}}\ket{\pi _{\rm tot}{} ^{  \alpha}{}_{\beta}} \ \ \ \ket{\psi} \in \mathcal{H}  \\ 
\ket{\pi_{\rm tot}{}^{\alpha}{}_{\beta}} &=& \ket{\pi ^{\alpha} {}_{\alpha _1}} \otimes \ket{\pi^{\alpha _1} {}_{\alpha _2} } \otimes \cdots \otimes \ket{\pi ^{\alpha _{N-1}}{}_{\beta}}
\ea
The value at $\ket{U_1, \cdots, U_N}$ becomes as follows.
\ba
\braket{U_1 , \cdots, U_N | \pi _{\rm tot}{}^{\alpha} {}_{\beta} } &=& \pi ^{\dagger} (U_1) _{\alpha_1}{}^{\alpha } \pi ^{\dagger} (U_2) _{\alpha_2}{}^{\alpha _1} \cdots \pi ^{\dagger} (U_N) _{\beta}{}^{\alpha _{N -1} } \notag \\
&=& \pi ^{\dagger} (U_N \cdots U_1) _{\beta} {}^{\alpha}.
\ea
This confirms that the physical Hilbert space is spanned by the functions on the group $G$.

For example,  we consider $V $ is an interval in the middle. In this case, the basis is given by
\be
\sqrt{\dim V_{\pi}} \ket{\pi_{\rm tot}{} ^{\alpha}{}_{\beta}} = (\dim V_{\pi}) ^{-1} \sum_{\gamma , \delta}\sqrt{\dim V_{\pi}} \ket{\pi_{\bar{V}} ^{\alpha} {}_{\gamma}} \otimes \sqrt{\dim V_{\pi}} \ket{\pi _V ^{\gamma} {}_{\delta}} \otimes\sqrt{\dim V_{\pi}} \ket{\pi_{\bar{V}} ^{\delta}{}_{\beta}} .
\ee 
From the decomposition, we find the reduced density matrix is given by 
\be
\rho _V = \sum_{\pi} p(\pi) (\dim V_{\pi}) ^{-2} \sum_{\gamma , \delta} (\sqrt{\dim V_{\pi}} \ket{\pi_{V} ^{\gamma} {}_{\delta}}) (\sqrt{\dim V_{\pi}} \bra{\pi_V ^{\delta }{}_{\gamma}})
\ee
where $p(\pi) = \sum \psi(\pi)_{\alpha} {}^{\beta} \psi^{*}(\pi) _{\beta} {}^{\alpha} $.
The expression of the reduced density matrix is the same with the case of periodic boundary condition (\ref{rdm}), so that the entanglement entropy is given by the same formula (\ref{eenonabg}) .

The topological state with open boundary condition is given by 
\be
\ket{\text{topo}}  = \frac{1}{\sqrt{|G|^{2N -1}}}\sum_{g _i \in G}\ket{g_1 ^{-1}}   \otimes \ket{g_2 ^{-1 } g _1} \otimes \cdots \otimes \ket{g_{N-1}^{-1}g_{N-2}} \otimes \ket{g _{N-1}}.
\ee
Thus $\sqrt{\dim V_{\pi}}\ket{\pi_{\rm tot}{} ^{\alpha }{}_{\beta}}$ component is obtained as
\be
\psi _{\alpha}{}^{\beta} (\pi) = \sqrt{\dim V_\pi} \braket{\pi ^{\beta}{}_{\alpha} | \text{topo}} = \sqrt{\frac{\dim V_{\pi}}{|G|}} \delta _{\alpha} {}^{\beta} ,
\ee
and $p(\pi)$ becomes
\be
p(\pi) = \frac{(\dim V_{\pi})^2}{|G|} ,
\ee
which is identical to the result with the periodic boundary condition case. We thus obtain the same result also for  the entanglement entropy as
\be
S_V = \log |G| .
\ee

\providecommand{\href}[2]{#2}\begingroup\raggedright

\endgroup

\end{document}